\documentclass[twocolumn]{aastex631}
\usepackage{amsmath}
\usepackage{multirow}
\usepackage{url}
\usepackage{color}
\usepackage{hyperref}
\usepackage{fnbreak}
\usepackage{subfigure}
\usepackage{acronym}
\usepackage{xspace}

\newcommand{\fref}[1]{Fig.~\ref{#1}}
\newcommand{\tref}[1]{Table~\ref{#1}}

\newcommand{\Tref}[1]{Table~\ref{#1}}

\newcommand{\mc}[1]{\mathcal{#1}}
\newcommand{\F}{\mc{F}}

\newcommand{\Tasc}{t_{\text{asc}}}
\newcommand{\TascThen}{\Tasc} \newcommand{\TascNow}{\Tasc'} \newcommand{\norb}{n_{\text{orb}}}

\newcommand{\Tmax}{T_{\text{max}}}

\newcommand{\Porb}{P_{\text{orb}}}
\newcommand{\PorbShear}{\tilde{P}}

\newcommand{\un}[1]{\,\text{#1}}
\newcommand{\code}[1]{\texttt{#1}\xspace}

\DeclareMathOperator{\Corr}{Corr}

\newcommand{\coord}{coordinate}

\newcommand{\PEGSi}{PEGS~I}
\newcommand{\PEGSii}{PEGS~II}
\newcommand{\PEGSiii}{PEGS~III}
\newcommand{\PEGSiv}{PEGS~IV}
\newcommand{\dcc}{LIGO-P2300042-v7}

\newcommand{\WangPorbSec}{68023.86}
\newcommand{\WangdPorbSec}{0.043}
\newcommand{\WangTascGPS}{974416624}
\newcommand{\WangTascDayTime}{2010--Nov--21 23:16:49\,UTC}

\newcommand{\WangdTascGPS}{50}
\newcommand{\OiiiTascGPS}{1255015049}
\newcommand{\OiiiTascDayTime}{2019--Oct--13 15:17:11\,UTC}

\newcommand{\OiiidTascGPS}{185}
\newcommand{\OiiiCorrCoeff}{0.96}
\newcommand{\KillPorbSec}{68023.91}
\newcommand{\KilldPorbSec}{0.017}
\newcommand{\KillTascGPS}{1078153676}
\newcommand{\KillTascDayTime}{2014--Mar--06 15:07:40\,UTC}

\newcommand{\KilldTascGPS}{33}
\newcommand{\newOiiiTascGPS}{1255015866}
\newcommand{\newOiiiTascDayTime}{2019--Oct--13 15:30:48\,UTC}

\newcommand{\newOiiidTascGPS}{55}
\newcommand{\newOiiiCorrCoeff}{0.80}

\newcommand{\newdPtildeSec}{0.010}
\newcommand{\newdPorbdTascShear}{2.47\times 10^{-4}}
\newcommand{\WangAsiniMin}{1.44}
\newcommand{\WangAsiniMax}{3.25}

\newcommand{\OiiiaStartGPS}{1238166018}

\newcommand{\OiiiaStartDay}{2019--Apr--01}

\newcommand{\OiiibEndGPS}{1269363618}

\newcommand{\OiiibEndDay}{2020--Mar--27}
\newcommand{\TmaxMin}{240}
\newcommand{\TmaxMax}{18720}
\newcommand{\onesigpct}{39.3}
\newcommand{\twosigpct}{86.5}
\newcommand{\threesigpct}{98.9}
\newcommand{\OiiiNorb}{4125}
\newcommand{\newOiiiNorb}{2600}
\newcommand{\mismatchMax}{0.25}

\newcommand{\numUnvetoed}{24}

\newcommand{\outlierf}{510.71}
\newcommand{\outlierrhoo}{6.08}
\newcommand{\outlierrhoi}{6.48}
\newcommand{\outlierrhoii}{9.58}
\newcommand{\outlierrhoiii}{10.41}
\newcommand{\outlierratiotwoone}{1.48}
\newcommand{\outlierratiotretwo}{1.09}
\begin{document}
\reportnum{\dcc}
\title[Sco~X-1 Search in LIGO O3 With Corrected Ephemeris]
{Search for Gravitational Waves from Scorpius~X-1 in LIGO O3 Data\\ With Corrected Orbital Ephemeris}
\author[0000-0001-5710-6576]{John~T.~Whelan}
\affiliation{School of Mathematical Sciences and Center for Computational Relativity and Gravitation, Rochester Institute of Technology, Rochester, NY 14623, USA}
\author[0000-0002-3582-2587]{Rodrigo~Tenorio}
\affiliation{Departament de F\'isica, Institut d'Aplicacions Computacionals i de Codi Comunitari (IAC3), Universitat de les Illes Balears, and Institut d'Estudis Espacials de Catalunya (IEEC), Carretera de Valldemossa km~7.5, E-07122 Palma, Spain}
\author[0000-0002-4301-2859]{Jared~K.~Wofford}
\affiliation{School of Physics and Astronomy and Center for Computational Relativity and Gravitation, Rochester Institute of Technology, Rochester, NY 14623, USA}
\author[0000-0003-3243-1393]{James~A.~Clark}
\affiliation{LIGO Laboratory, California Institute of Technology, Pasadena, CA 91125, USA}
\author[0000-0002-3780-5430]{Edward~J.~Daw}
\affiliation{The University of Sheffield, Sheffield S10 2TN, United Kingdom}
\author[0000-0003-2666-721X]{Evan~Goetz}
\affiliation{University of British Columbia, Vancouver, BC V6T 1Z4, Canada}
\author[0000-0002-2824-626X]{David~Keitel}
\affiliation{Departament de F\'isica, Institut d'Aplicacions Computacionals i de Codi Comunitari (IAC3), Universitat de les Illes Balears, and Institut d'Estudis Espacials de Catalunya (IEEC), Carretera de Valldemossa km~7.5, E-07122 Palma, Spain}
\author[0000-0003-0323-0111]{Ansel~Neunzert}
\affiliation{LIGO Hanford Observatory, Richland, WA 99352, USA}
\author[0000-0001-9050-7515]{Alicia~M.~Sintes}
\affiliation{Departament de F\'isica, Institut d'Aplicacions Computacionals i de Codi Comunitari (IAC3), Universitat de les Illes Balears, and Institut d'Estudis Espacials de Catalunya (IEEC), Carretera de Valldemossa km~7.5, E-07122 Palma, Spain}
\author[0000-0002-7255-4251]{Katelyn~J.~Wagner}
\affiliation{School of Physics and Astronomy and Center for Computational Relativity and Gravitation, Rochester Institute of Technology, Rochester, NY 14623, USA}
\author[0000-0003-0381-0394]{Graham~Woan}
\affiliation{SUPA, University of Glasgow, Glasgow G12 8QQ, United Kingdom}
\author[0000-0002-0440-9597]{Thomas~L.~Killestein}
\affiliation{Department of Physics, University of Warwick, Gibbet Hill Road, Coventry CV4 7AL, United Kingdom}
\author[0000-0003-0771-4746]{Danny~Steeghs}
\affiliation{Department of Physics, University of Warwick, Gibbet Hill Road, Coventry CV4 7AL, United Kingdom}
\affiliation{OzGRav-Monash, School of Physics and Astronomy, Monash University, Victoria 3800, Australia}
\date{2023 February 20}
\begin{abstract}
  Improved observational constraints on the orbital parameters of the
  low-mass X-ray binary Scorpius~X-1 were recently published
  in \cite{Killestein2023_PEGS4}. In the
  process, errors were corrected in previous orbital ephemerides,
  which have been used in searches for continuous gravitational waves
  from Sco~X-1 using data from the Advanced LIGO detectors.  We
  present the results of a re-analysis of LIGO detector data from the
  third observing run of Advanced LIGO and Advanced Virgo using a
  model-based cross-correlation search.
  The corrected region of parameter space, which
  was not covered by previous searches, was about 1/3 as large as
  the region searched in the original O3 analysis, reducing the required
  computing time.
  We have confirmed that no detectable signal is present over a range of
  gravitational-wave frequencies from $25\un{Hz}$ to $1600\un{Hz}$,
  analogous to the null result of \cite{LVK2022_O3ScoX1CrossCorr}.  Our search
  sensitivity is comparable to that of \cite{LVK2022_O3ScoX1CrossCorr},
  who set upper limits
  corresponding, between $100\un{Hz}$ and $200\un{Hz}$, to
  an amplitude $h_0$ of about $10^{-25}$ when marginalized
  isotropically over the unknown inclination angle of the neutron
  star's rotation axis, or less than $4\times 10^{-26}$ assuming the
  optimal orientation.
\end{abstract}

\acrodef{NS}[NS]{neutron star}
\acrodef{GW}[GW]{gravitational wave}
\acrodef{LMXB}[LMXB]{low-mass X-ray binary}
\acrodef{BBH}[BBH]{binary black hole}
\acrodefplural{LMXB}[LMXBs]{low-mass X-ray binaries}
\acrodef{AMXP}[AMXP]{accreting millisecond X-ray pulsar}
\acrodef{EM}[EM]{Electromagnetic}
\acrodef{CW}[CW]{continuous wave}
\acrodef{PSD}[PSD]{power spectral density}
\acrodef{SNR}[SNR]{signal-to-noise ratio}
\acrodef{CPU}[CPU]{central processing unit}
\acrodef{SFT}[SFT]{short Fourier transform}
\acrodef{LLO}[LLO]{LIGO Livingston Observatory}
\acrodef{LHO}[LHO]{LIGO Hanford Observatory}
\acrodef{IFO}[IFO]{interferometer}
\acrodef{GPS}[GPS]{Global Positioning System}
\acrodef{ScoX1}[Sco~X-1]{Scorpius~X-1}
\acrodef{EoS}[EoS]{equation of state}
\acrodefplural{EoS}{equations of state}
\preprint{\dcc}
\section{Introduction}
\label{s:intro}

The \ac{LMXB} \ac{ScoX1}, which is presumed to consist of a \ac{NS} of
mass $\approx1.4M_{\odot}$ in a binary orbit with a companion star of
mass $\approx0.4M_{\odot}$ \citep{Steeghs2002_ScoX1}, is a very
promising potential source of continuous \acp{GW}, generated by the
spin of the \ac{NS} \citep{Bildsten1998,Watts2008}.  As such, it has
been the target of a number of searches
\citep{LSC2007_S2ScoX1FStat,LVC2015_S5ScoX1Sideband,Meadors2017_S6ScoX1TwoSpect,LVC2017_O1StochRadiometer,LVC2017_O1ScoX1Viterbi,LVC2017_O1ScoX1CrossCorr,LVC2019_O2StochRadiometer,LVC2019_O2ScoX1Viterbi,Zhang2021_O2ScoX1CrossCorr,LVK2021_O3StochRadiometer,LVK2022_O3ScoX1Viterbi,LVK2022_O3ScoX1CrossCorr}
using data from the Advanced LIGO \ac{GW} detectors
\citep{LVC2015_aLIGOInst}, which have conducted three observing runs
(O1, O2 and O3), the last two in coordination with Advanced Virgo
\citep{Acernese2015_aVirgo}.  As the spin frequency of \ac{ScoX1} is
unknown, searches typically cover a wide range of intrinsic \ac{GW}
signal frequency $f_0$, which in the simplest model is twice the spin
frequency.  Some of these searches, notably the
Cross-Correlation method
\citep{Dhurandhar2007_CrossCorr,Whelan2015_ScoX1CrossCorr} and the
Viterbi method \citep{Suvorova2016_Viterbi,Suvorova2017_Viterbi2}, are sensitive to aspects
of the signal model, notably the parameters of the binary orbit. Hence, a
campaign of electromagnetic observations and analyses known as
``Precision Ephemerides for Gravitational-Wave Searches'' (PEGS) has
produced a series of updates to the orbital ephemeris for \ac{ScoX1}
\citep{Galloway2014_PEGS1,Wang2018_PEGS3,Killestein2023_PEGS4} which
have been used to choose the parameter space region covered in the
\ac{GW} searches.  The most recent update, known as
{\PEGSiv} \citep{Killestein2023_PEGS4}, in addition to producing a
more refined ephemeris, also corrected errors in the previously
published {\PEGSi} \citep{Galloway2014_PEGS1} and {\PEGSiii}
\citep{Wang2018_PEGS3} ephemerides.\footnote{{\PEGSii}
  \citep{Premachandra2016_PEGS2} was an ephemeris for Cygnus X-2
  rather than \ac{ScoX1}, and is not relevant to this work.}  This means
that the parameter space region searched by the analysis of O1 data in
\cite{LVC2017_O1ScoX1CrossCorr}, which used elements of {\PEGSi} and
{\PEGSiii}, as well as the O2 and O3 analyses in
\cite{LVC2019_O2ScoX1Viterbi,Zhang2021_O2ScoX1CrossCorr,LVK2022_O3ScoX1Viterbi,LVK2022_O3ScoX1CrossCorr},
which used {\PEGSiii}, did not overlap with the likely regions of
parameter space according to the {\PEGSiv} ephemeris, nor with the
revised {\PEGSi} and {\PEGSiii} ephemerides published in
\cite{Killestein2023_PEGS4}.\footnote{The analysis of LIGO O1 data in
  \cite{LVC2017_O1ScoX1Viterbi} used the orbital period from the
  {\PEGSi} ephemeris, but since it used the method of
  \cite{Suvorova2016_Viterbi,Suvorova2017_Viterbi2}, which does not require the orbital
  phase, it was not affected in the same way as the other analyses.}
This paper presents a re-analysis of the
LIGO O3 data \citep{LVK2023_O3OpenData} according to the method of
\cite{LVK2022_O3ScoX1CrossCorr}, but with the parameter space
determined by the {\PEGSiv} ephemeris.

\begin{deluxetable}{l c c}
  \tablewidth{\textwidth}
  \tablecaption{Orbital parameters in the {\PEGSiii} and {\PEGSiv} ephemerides.
    \label{t:ScoX1ephem}}
  \tablehead{
    \colhead{Parameter} & \colhead{{\PEGSiii}\tablenotemark{a}}
    & \colhead{{\PEGSiv}\tablenotemark{b}}
  }
  \startdata
  $\Porb$ (s) & $\WangPorbSec\pm\WangdPorbSec$
  & $\KillPorbSec\pm\KilldPorbSec$
  \\
  $\TascThen$ (GPS s)\tablenotemark{c} & $\WangTascGPS\pm\WangdTascGPS$
  & $\KillTascGPS\pm\KilldTascGPS$
  \\
  $\TascNow$ (GPS s)\tablenotemark{d} & $\OiiiTascGPS\pm\OiiidTascGPS$
  & $\newOiiiTascGPS\pm\newOiiidTascGPS$
  \\
  $\Corr(\Porb,\TascNow)$ \tablenotemark{e} & $\OiiiCorrCoeff$
  & $\newOiiiCorrCoeff$
  \enddata
  \tablerefs{\cite{Wang2018_PEGS3,Killestein2023_PEGS4}}
\tablecomments{Uncertainties are $1\sigma$.}
  \tablenotetext{a}{Values in this column are inferred from
    \cite{Wang2018_PEGS3}}
\tablenotetext{b}{Values in this column are inferred from
    \cite{Killestein2023_PEGS4}}
\tablenotetext{c}{The time of ascension $\TascThen$, at which the
    \ac{NS} crosses the ascending node (moving away from the
    observer), measured in the Solar System barycenter, is derived
    from the time of inferior conjunction of the companion by
    subtracting $\Porb/4$.  The values quoted in this row are those
    for which the correlations in the $\Porb$ and $\TascThen$
    uncertainties are negligible, and correspond to
    {\WangTascDayTime} and {\KillTascDayTime}, respectively.}
\tablenotetext{d}{The time of ascension $\TascNow$ after propagating
    $\TascThen$ forward by $\norb$ orbits ({\OiiiNorb} and
    {\newOiiiNorb}, respectively), corresponding to times of
    {\OiiiTascDayTime} and {\newOiiiTascDayTime}, near the middle of
    the O3 run.  The uncertainty is obtained by combining the
    uncertainty in $\TascNow$ in quadrature with $\norb$ times the
    uncertainty in $\Porb$ \citep{Whelan2015_ScoX1CrossCorr}.}
\tablenotetext{e}{The correlation between the uncertainties in
    $\Porb$ and $\TascNow=\TascThen+\norb\Porb$, given uncorrelated
    uncertainties in $\Porb$ and $\TascThen$.}
\end{deluxetable}

\begin{figure}[tbp]
  \centering
  \includegraphics[width=\columnwidth]{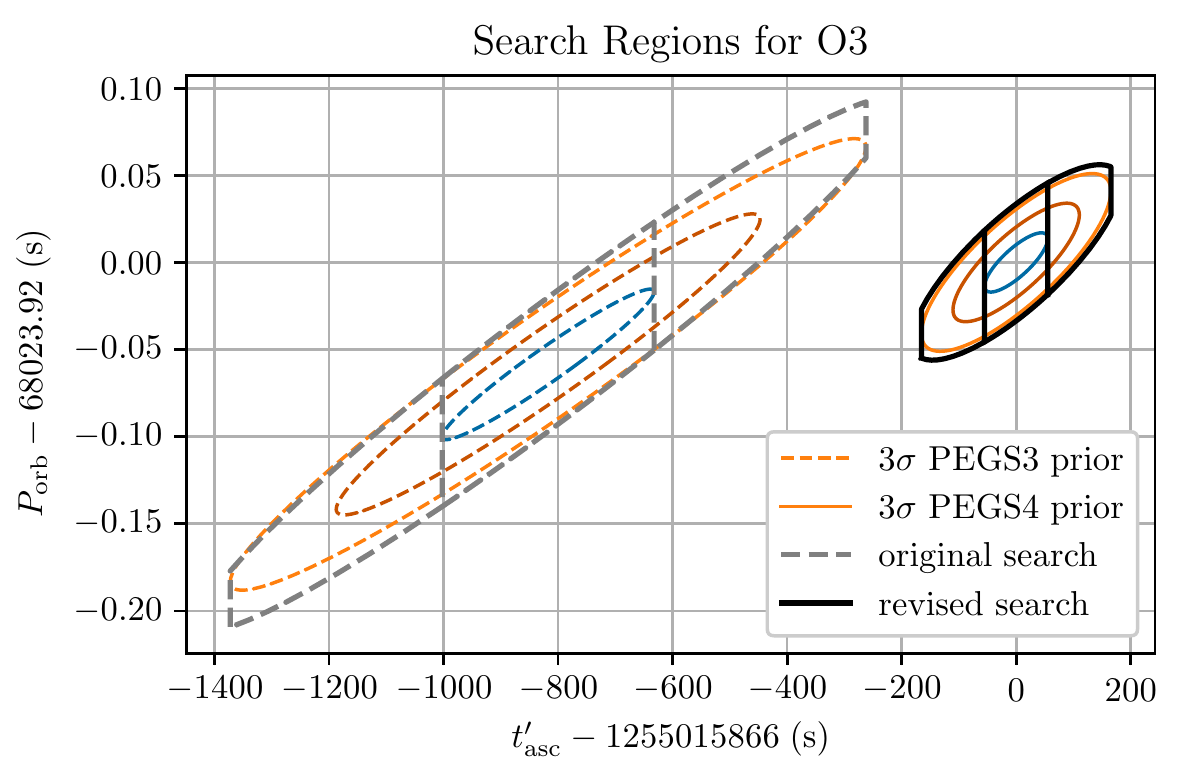}
  \caption{Prior uncertainties and search regions for O3 analyses.
    The dashed colored ellipses show curves of constant prior
    probability according to the {\PEGSiii} ephemeris of
    \cite{Wang2018_PEGS3}, corresponding to $1\sigma$, $2\sigma$, and
    $3\sigma$ (containing $\onesigpct\%$, $\twosigpct\%$, and
    $\threesigpct\%$ of the prior probability, respectively).  The
    solid colored ellipses show the same curves for the {\PEGSiv}
    ephemeris of \cite{Killestein2023_PEGS4}.  The dashed gray lines
    show the search regions used in the O3 Cross-Correlation analysis
    of \cite{LVK2022_O3ScoX1CrossCorr}; the central region, between
    the vertical lines, was searched with higher coherence times.  The
    solid black lines show the search region for the present
    re-analysis.}
  \label{f:TPRegions}
\end{figure}

\section{Orbital Ephemerides for Scorpius~X-1}

The signal model for \acp{GW} from an \ac{LMXB} is a continuous signal
with nearly constant intrinsic amplitude and frequency,
Doppler-modulated by the relative motion of the source and the
detector.  As the Doppler modulation depends upon the extrinsic
parameters of the system, including the sky position and orbital
parameters of the binary system, accurate ranges of values for those
parameters are an important input into \ac{GW} searches.  Since the
sky location of \ac{ScoX1} is precisely known
\citep{Bradshaw1999_ScoX1,LSC2007_S2ScoX1FStat}, and the orbital eccentricity
is believe to be small
\citep{Steeghs2002_ScoX1,Wang2018_PEGS3,Killestein2023_PEGS4}, the
important residual uncertainty is in the orbital velocity, period, and
phase of the \ac{NS}.  The projected orbital velocity $K_1$ of the
\ac{NS} is usually described for \ac{GW} searches in terms of the
projected semimajor axis $a\sin i=K_1\Porb/(2\pi)$ of the orbit,
measured in light-seconds.\footnote{Since the fractional uncertainty
  on $\Porb$ is much smaller than that in $K_1$, no significant
  correlation between $a\sin i$ and $\Porb$ is introduced by this
  convention.}  Estimation of $K_1$ is particularly difficult for
\ac{ScoX1} \citep{Galloway2014_PEGS1}, and the best constraint remains
that of \cite{Wang2018_PEGS3},
$40\un{km/s}\lesssim K_1\lesssim 90\un{km/s}$.  The orbital phase is
generally described by the time at which the system reaches some
reference point in its orbit.  For \ac{GW} searches, this is typically
the time of ascension $\Tasc$, when the \ac{NS} crosses the ascending
node, moving away from the observer.  This is one-quarter period
before the other typically-quoted reference time, of inferior
conjunction of the companion star.  Given a value of $\TascThen$ and
$\Porb$, an equivalent time of ascension in a later epoch
$\TascNow=\TascThen + \norb\Porb$ can be obtained by adding an integer
number $\norb$ of orbits \citep{Whelan2015_ScoX1CrossCorr}.
For analysis of LIGO O3 data, it is
convenient to choose a $\TascNow$ value in the middle of the observing
run, which lasted from {\OiiiaStartDay} to {\OiiibEndDay},
corresponding to GPS time {\OiiiaStartGPS} to {\OiiibEndGPS}.

The parameter space ranges used for searches of O2 and O3 data
\citep{LVC2019_O2ScoX1Viterbi,Zhang2021_O2ScoX1CrossCorr,LVK2022_O3ScoX1Viterbi,LVK2022_O3ScoX1CrossCorr}
were generated using the {\PEGSiii} ephemeris \citep{Wang2018_PEGS3}.
\cite{Killestein2023_PEGS4} subsequently published the improved {\PEGSiv}
ephemeris, also documenting calibration errors in {\PEGSiii}.  The
values of $\Porb$ and $\TascThen$ in these ephemerides are summarized
in \tref{t:ScoX1ephem}, which shows that the $\Tasc$ range originally
published in \cite{Wang2018_PEGS3} is inconsistent with current
estimates of that parameter.
In \fref{f:TPRegions} we show the plausible ranges of $\Porb$ and
$\TascNow$ (propagated to the middle of O3), along with the region of
parameter space searched in the Cross-Correlation analysis
\citep{LVK2022_O3ScoX1CrossCorr}, and in the re-analysis presented in
this paper.  We see that the parameter space region searched in the
original O3 search is inconsistent with the {\PEGSiv} ephemeris, but
the region searched in the present re-analysis has about 1/3 the area
in parameter space, allowing a search to be performed more quickly at
the same sensitivity.

\begin{figure}[tbp]
  \centering
  \includegraphics[width=\columnwidth]{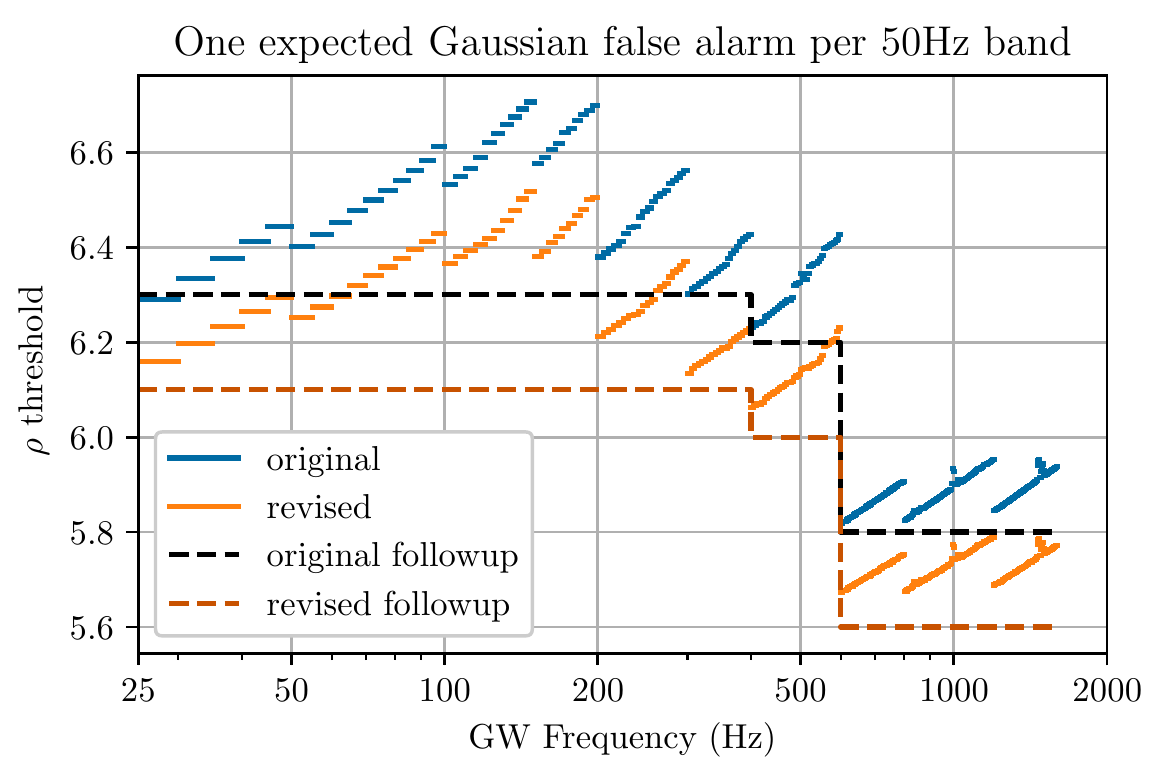}
  \caption{Selection of followup threshold as a function of \ac{GW}
    frequency.  If the data contained no signal and only Gaussian
    noise, each template in the parameter space would have some chance
    of producing a statistic value exceeding a given threshold.
    Within each $5\un{Hz}$ frequency band, the total number of
    templates was computed and used to find the threshold at which the
    expected number of Gaussian outliers (assuming uncorrelated templates)
    above that value would be
    $0.1$.  The short blue lines show this quantity for the
    original O3 search in \cite{LVK2022_O3ScoX1CrossCorr}; the short
    green lines show this for the present re-analysis.  Because of the
    smaller parameter space searched with the {\PEGSiv} ephemeris, the
    present search uses fewer templates (\textit{cf.}\
    \tref{t:SearchSummary}), and therefore would be expected to have
    the specified number of false alarms at a lower threshold.  We
    thus use a lower threshold for follow-ups in this analysis (dashed
    red line) than in the original analysis (dashed black line.)
    Compare Fig.~4 of \cite{LVK2022_O3ScoX1CrossCorr}.}
\label{f:threshold}
\end{figure}

\begin{figure}[tbp]
  \centering
  \includegraphics[width=\columnwidth]{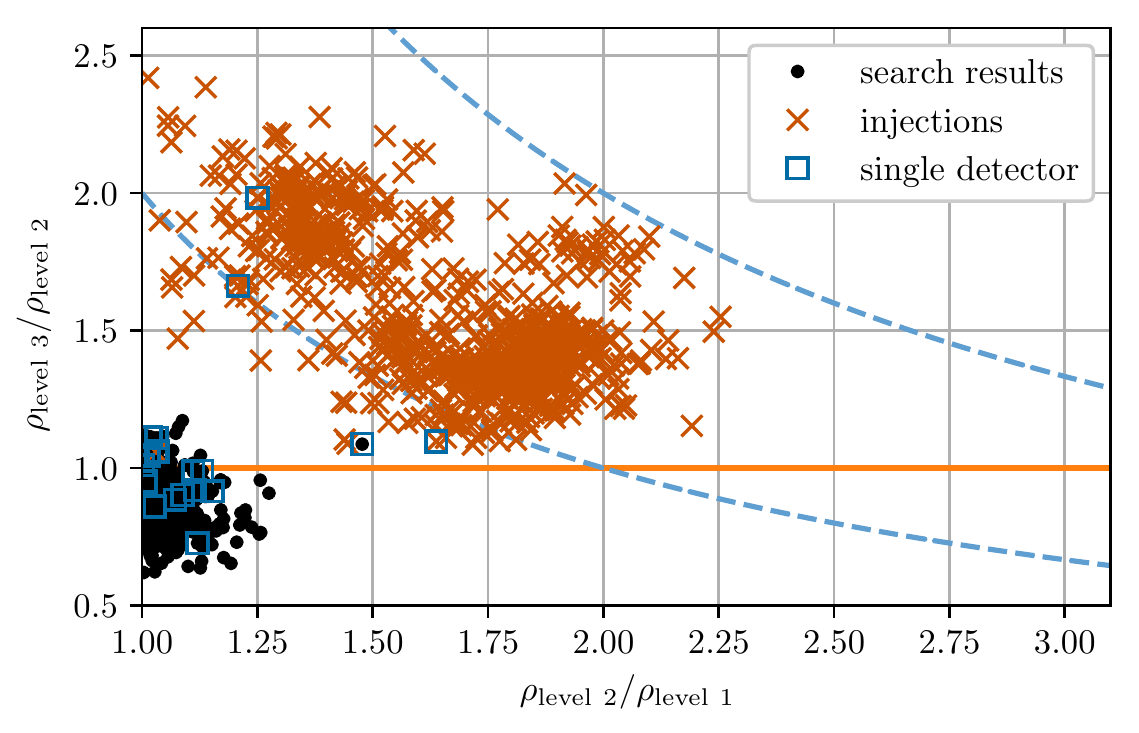}
  \caption{Ratios of followup statistics for search candidates and
    simulated signals.  This plot shows all of the candidates that
    survived level~2 of followup (see \tref{t:SearchSummary}).  It
    shows the ratios of the \ac{SNR} $\rho$ after followup level~1 (at
    the original coherence time $\Tmax$), level~2 (at $4\times$ the
    original coherence time), and level~3 (at $16\times$ the original
    coherence time).  For comparison, the results of the original
    injection analysis from \cite{LVK2022_O3ScoX1CrossCorr} are shown.
    The dashed lines are at constant values of
    $\rho_{\text{level 3}}/\rho_{\text{level 1}}$ equal to $2$ and
    $4$.  Points below the solid line have
    $\rho_{\text{level 3}}<\rho_{\text{level 2}}$ and are therefore
    vetoed at followup level 2.  The boxes labelled ``single
    detector'' are outliers or injections at \ac{GW} frequencies where
    only one detector's data was included in the analysis because of
    known instrumental artifacts in the other detector.  Compared to
    Figure~5 of \cite{LVK2022_O3ScoX1CrossCorr}, we see one candidate
    (with
    $(\frac{\rho_{\text{level 2}}}{\rho_{\text{level 1}}},
    \frac{\rho_{\text{level 3}}}{\rho_{\text{level 2}}}) \approx
    (\outlierratiotwoone,\outlierratiotretwo)$) which increases its
    \ac{SNR} marginally similarly to the least significant of the
    injections.  As discussed in the text, this is a ``single
    detector'' outlier where only \ac{LHO} data have been used, and it
    appears to be an instrumental artifact.}
\label{f:followupRatio}
\end{figure}

\begin{deluxetable*}{rrrrclcrrrr}
  \tablewidth{0.9\textwidth}
  \tablecaption{Summary of Numbers of Templates and Candidates.
    \label{t:SearchSummary}}
  \tablehead{
\multicolumn{2}{c}{$f_0$(Hz)} &
\multicolumn{2}{c}{$\Tmax$(s)} &
\colhead{$\rho$} &
\colhead{number of} &
\colhead{expected Gauss.} &
\multicolumn{4}{c}{followup level} \\
\colhead{min} &
\colhead{max} &
\colhead{min} &
\colhead{max} &
\colhead{thresh\tablenotemark{a}} &
\colhead{templates} &
\colhead{false alarms\tablenotemark{b}} &
\colhead{0\tablenotemark{c}} &
\colhead{1\tablenotemark{d}} &
\colhead{2\tablenotemark{e}} &
\colhead{3\tablenotemark{f}}
}
\startdata
25 &
50 &
10080 &
18720 &
6.1 &
$2.24\times 10^{9}$ &
1.2 &
62 &
34 &
5 &
0
\\
50 &
100 &
8160 &
14280 &
6.1 &
$9.65\times 10^{9}$ &
5.1 &
134 &
115 &
41 &
0
\\
100 &
150 &
6720 &
10920 &
6.1 &
$1.71\times 10^{10}$ &
9.1 &
148 &
146 &
81 &
3
\\
150 &
200 &
5040 &
8640 &
6.1 &
$1.79\times 10^{10}$ &
9.5 &
151 &
151 &
74 &
1
\\
200 &
300 &
2400 &
4800 &
6.1 &
$1.30\times 10^{10}$ &
6.9 &
60 &
60 &
22 &
8
\\
300 &
400 &
1530 &
3060 &
6.1 &
$6.51\times 10^{9}$ &
3.5 &
24 &
24 &
5 &
3
\\
400 &
600 &
720 &
2160 &
6.0 &
$9.96\times 10^{9}$ &
9.8 &
312 &
199 &
20 &
9
\\
600 &
800 &
360 &
360 &
5.6 &
$7.34\times 10^{8}$ &
7.9 &
8 &
8 &
0 &
0
\\
800 &
1200 &
300 &
300 &
5.6 &
$1.69\times 10^{9}$ &
18.2 &
234 &
65 &
2 &
0
\\
1200 &
1600 &
240 &
240 &
5.6 &
$1.65\times 10^{9}$ &
17.7 &
328 &
61 &
7 &
0
\\
\enddata
 \tablecomments{For each range of \ac{GW} frequencies, this table
    shows the minimum and maximum coherence time $\Tmax$ used for the
    search, the threshold in the \ac{SNR} $\rho$ used for followup, the
    total number of templates, and the number of candidates at various
    stages of the process.  Compare Table~3 of
    \cite{LVK2022_O3ScoX1CrossCorr}.}
\tablenotetext{a}{This is the threshold for initiating followup,
    i.e., to produce a level 0 candidate.}
\tablenotetext{b}{This is the number of candidates that would be
    expected in Gaussian noise, given the number of templates and the
    followup threshold.}
\tablenotetext{c}{This is the actual number of candidates (after
    clustering) which crossed the \ac{SNR} threshold and were followed
    up.}
\tablenotetext{d}{This is the number of candidates remaining after
    refinement.  All of the candidates ``missing'' at this stage have
    been removed by the single-detector veto for unknown lines.}
\tablenotetext{e}{This is the number of candidates remaining after
    each has been followed up with a $\Tmax$ equal to $4\times$ the
    original $\Tmax$ for that candidate.  (True
    signals should approximately double their \ac{SNR}; any candidates
    whose \ac{SNR} goes down have been dropped.)  All of the signals
    present at this stage are shown in \fref{f:followupRatio}, which also
    shows the behavior of the search on simulated signals injected in
    software.}
\tablenotetext{f}{This is the number of candidates remaining after
    $\Tmax$ has been increased to $16\times$ its original value.}
\vspace{10pt}
\end{deluxetable*}

\section{Re-Analysis of LIGO O3 Data with Cross-Correlation Pipeline}

We present here the results of a re-analysis of the LIGO O3 data
\citep{LVK2023_O3OpenData} using
the Cross-Correlation search \citep{Whelan2015_ScoX1CrossCorr}, with
the revised {\PEGSiv} ephemeris of \cite{Killestein2023_PEGS4}.  Full
details of the analysis pipeline are given in
\cite{LVK2022_O3ScoX1CrossCorr}.  We highlight here changes made in
light of the revised ephemeris.

The search was performed over a range of signal frequencies from $25$
to $1600\un{Hz}$.  Since the search is tunable, with the coherence
time $\Tmax$ chosen to balance computing cost and sensitivity,
different $\Tmax$ values are chosen across signal frequency and
orbital parameter space to roughly optimize the chance of detecting a
signal.  The same coherence times were used as in
\cite{LVK2022_O3ScoX1CrossCorr}, with
$\TmaxMin\un{s}\le\Tmax\le\TmaxMax\un{s}$.  As in
\cite{LVK2022_O3ScoX1CrossCorr} the search covered a range of
projected semimajor axes
$\WangAsiniMin\un{lt-s}\le a\sin i\le\WangAsiniMax\un{lt-s}$.\footnote{There
  was a slight difference arising from converting the range
  $40\le K_1\le 90\un{km/s}$ to $a\sin i$ using the revised period
  estimate, but well below the precision reported here.}  The
$\TascNow$-$\Porb$ space was covered using the sheared period
coordinate of \cite{Wagner2022_Lattice}, which for the re-analysis was
defined as
\begin{equation}
  \label{e:Ptildedef}
  \PorbShear = \Porb - \newdPorbdTascShear (\TascNow-\newOiiiTascGPS)
  \ .
\end{equation}
The range of $\TascNow$ values was set to
$\newOiiiTascGPS\pm3\times\newOiiidTascGPS$, while the $\Porb$ values
were constrained to lie in an ellipse centered on
$(\PorbShear,\TascNow)=(\KillPorbSec\un{s},\newOiiiTascGPS)$ with
semiaxes of $3.3\times \newdPtildeSec\un{s}$ for $\PorbShear$ and
$3.3\times \newOiiidTascGPS\un{s}$ for $\TascNow$.  The boundaries of
this region in $(\Porb,\TascNow)$ space are shown in solid black lines
in \fref{f:TPRegions}.  Note that the software bug which led to the
slightly misaligned definition of $\PorbShear$ used in
\cite{LVK2022_O3ScoX1CrossCorr} was fixed before the re-analysis, so
the search region now lines up with the {\PEGSiv} prior uncertainty
region, as seen in \fref{f:TPRegions}.  As in
\cite{LVK2022_O3ScoX1CrossCorr}, the $\PorbShear$ {\coord} was
unresolved for most analysis jobs and a single template was sufficient to cover
the range $\KillPorbSec\pm 3.3\times\newdPtildeSec\un{s}$.

As in \cite{LVK2022_O3ScoX1CrossCorr} the search was carried out at a
nominal parameter-space mismatch of $\mismatchMax$.  Due to the
reduced parameter ranges in the more precise {\PEGSiv} ephemeris, this
required fewer templates than in the original analysis.  As a
consequence, the threshold for followup, which is set using the
expected number of false alarms from Gaussian noise at a particular
frequency, could be reduced, as shown in \fref{f:threshold}.
In the present search, we followed up candidates with an \acf{SNR}
above $6.1$ from $25\un{Hz}<f_0<400\un{Hz}$, $6.0$ for
$400\un{Hz}<f_0<600\un{Hz}$, and $5.6$ for
$600\un{Hz}<f_0<1600\un{Hz}$, compared to $6.3$, $6.2$, and $5.8$,
respectively, in the original analysis of
\cite{LVK2022_O3ScoX1CrossCorr}.

Detection candidates which exceeded the followup threshold were
subjected to a hierarchical followup using successively finer grids
and longer coherence times.  At each stage candidates for which the
\ac{SNR} from a search using only one detector (\acf{LHO} or \ac{LLO})
exceeded the \ac{SNR} obtained from the full data were rejected as
likely narrow-band instrumental features (``lines'').  The initial
results were known as ``level~0''; ``level~1'' used the same coherence
time $\Tmax$ and only refined the grid, while ``level~2'' and
``level~3'' each successively quadrupled $\Tmax$ relative to the
previous level, which would ideally double the \ac{SNR} of a signal.
If a candidate's \ac{SNR} went down from one level of followup to the
next, it was discarded.  \Tref{t:SearchSummary} shows the numbers of
candidates surviving each level of followup.  A total of
{\numUnvetoed} candidates survived level~3 of followup.  For each of
the candidates surviving level~2 (so that level~3 followup was run),
\fref{f:followupRatio} shows the ratios of \acp{SNR} at successive levels.

For the most part, we reproduce the results of
\cite{LVK2022_O3ScoX1CrossCorr}, that the outliers of the search do
not increase their SNR upon followup in the way that simulated signals
do.  There is one possible exception: a candidate at a frequency of
$\outlierf\un{Hz}$ which has \ac{SNR} $\rho$ of $\outlierrhoo$ at
level 0, $\outlierrhoi$ at level 1, $\outlierrhoii$ at level 2, and
$\outlierrhoiii$ at level 3.
Note that while this
candidate would not have made the followup threshold in the original
search, which was $6.2$ for frequencies between $400$ and $600\un{Hz}$
(as compared with 6.0 in the followup), there is no such outlier in
the original search with the {\PEGSiii} ephemeris; all of the
templates searched with $510.70\un{Hz}<f_0<510.72\un{Hz}$ produced
\ac{SNR} $\rho<4.5$ in the original search.

We have strong reasons to believe the outlier in the reanalysis is an instrumental
artifact. 
The outlier is located in a frequency range heavily contaminated by
violin modes in both LIGO detectors \citep{Davis2021_DetChar}.
The cross-correlation analysis excludes data at frequencies
contaminated by known lines \citep{Goetz2021_Lines} from the analysis.
For \ac{LHO} this
includes data from $510.71527$ to $510.72653\un{Hz}$, and for \ac{LLO}
from $507.89972$ to $516.25972\un{Hz}$.  The relative amplitude of possible
Doppler modulation for \ac{ScoX1} is $\lesssim 4\times 10^{-4}$, so a
signal with intrinsic frequency $510.7\un{Hz}$ could be received at
the detector with a frequency from $\sim 510.5\un{Hz}$ to
$\sim 510.9\un{Hz}$.  Thus \ac{LLO} is completely excluded from
analysis of this candidate, as indicated by the blue ``single
detector'' square in \fref{f:followupRatio}.  As a consequence, the
\ac{SNR} using only \ac{LHO} data is the same as the \ac{SNR} from the
full search, and the ``unknown line'' veto cannot be applied to this
candidate.  

As an additional investigation, we re-ran the followup
with no data excluded.  The \ac{SNR} of the candidate dropped to
$4.95$, while the \ac{SNR} using only \ac{LHO} increased to $65.90$.
(The \ac{SNR} using only \ac{LLO} was $2.45$.)  I.e., without the
``known line'' veto to eliminate \ac{LLO} data, the outlier would have
been eliminated as a possible candidate by the ``unknown line'' veto.

This outlier was further
scrutinized by a multistage MCMC follow-up using the method described
in~\citep{Tenorio2021_MCMC} with the PyFstat package~\citep{Ashton2018_PyFstat,
Keitel2021_PyFstat, ashton_gregory_2022_7458002}. The setup was identical
to that of~\citet{LVK2022_O3ScoX1CrossCorr}: Templates were placed
adaptively around the outlier to compute the semicoherent 
$\mathcal{F}$-statistic~\citep{JKS1998_FStat,Cutler2005_FStat} using
a decreasing number of coherent segments (660, 330, 92, 24, 4, and 1),
which correspond to a coherence time ranging from half a day to the full
observing run. A Bayes factor was computed using the $\F$-statistic values
of consecutive stages corresponding to the loudest template. The signal 
hypothesis assesses the consistency of these values, while comparing against the noise hypothesis
checks the inconsistency of the final value with the background distribution.
The resulting Bayes factor is significantly lower than expected for a signal
detectable by this search. Moreover, the semicoherent $\mathcal{F}$-statistic accumulation 
of the outlier in the \ac{LHO} data suggests an instrumental origin.

\section{Conclusions and Outlook}

We have re-analyzed the LIGO O3 data with the Cross-Correlation
pipeline, using the corrected and improved {\PEGSiv} ephemeris of
\cite{Killestein2023_PEGS4}.  Having found no credible detection
candidates, we reproduce the null result of \cite{LVK2022_O3ScoX1CrossCorr},
that there is no \ac{GW} signal from \ac{ScoX1} detectable at the
level of sensitivity of that search.  We do not produce an upper limit
from the re-analysis, but as approximate sensitivity computations for
the two searches agree, we conclude that the upper limits published
\cite{LVK2022_O3ScoX1CrossCorr} remain valid.

Since the Cross-Correlation analysis of this paper, as in
\cite{LVK2022_O3ScoX1CrossCorr}, does not explicitly consider a signal
with stochastically varying frequency (``spin wandering'')--although
it is somewhat robust to it \citep{Whelan2015_ScoX1CrossCorr}--there is
information to be gained from re-analysis of the O3 data using a
hidden Markov model as in \cite{LVK2022_O3ScoX1Viterbi}.

The O4 run of Advanced LIGO, Advanced Virgo and KAGRA
\citep{Akutsu2021_KAGRAO3} is scheduled to begin in May 2024 and run
for approximately 18 months \citep{LVK2023_ObsPlanUpdate}.
The improved sensitivity of the
detectors will enable more sensitive searches for \acp{GW} from
\ac{ScoX1}, and the greater precision of the {\PEGSiv} ephemeris will
enable the search to be done more efficiently.

\section*{Acknowledgments}

We wish to thank the members of the LIGO-Virgo-KAGRA Collaboration
continuous waves group for useful feedback.
JTW, JKW, and KJW were supported by NSF grants PHY-1806824 and PHY-2110460.
RT, DK, and AMS are supported by the Spanish Ministerio de Ciencia e
Innovaci\'on and the Spanish Agencia Estatal de Investigaci\'on grant
PID2019-106416GB-I00/AEI/MCIN/10.13039/501100011033, European Union
NextGenerationEU funds (PRTR-C17.I1), the Comunitat Aut\`onoma de les
Illes Balears through the Direcci\'o General de Pol\'itica
Universitaria i Recerca with funds from the Tourist Stay Tax Law ITS
2017-006 (PRD2018/24, PRD2020/11), the Conselleria de Fons Europeus,
Universitat i Cultura del Govern de les Illes Balears, the FEDER
Operational Program 2021--2027 of the Balearic Islands, and EU COST
Actions CA18108 and CA17137.
RT is supported by the Spanish Ministerio de Universidades (ref.~FPU 18/00694).
DK is supported by the Spanish Ministerio de Ciencia, Innovaci\'on y
Universidades (ref.~BEAGAL 18/00148) and cofinanced by the Universitat
de les Illes Balears.
TLK is supported by the UK Science and Technology Facilities Council
(STFC), grant number ST/T506503/1.
DS is supported by STFC, grant numbers ST/T007184/1, ST/T003103/1 and
ST/T000406/1.

The authors are grateful for computational resources provided by LIGO
Laboratory, supported by National Science Foundation Grants
PHY-1626190 and PHY-1700765, by the OSG Consortium
\citep{Pordes2007_OSG,Sfiligoi2009_OSG,OSGPool}, which is supported by
the National Science Foundation awards \#2030508 and \#1836650, and by
the Digital Research Alliance of Canada (\url{https://alliancecan.ca}).
Scripts and data pertaining to the original O3 analysis are available
at \url{https://dcc.ligo.org/LIGO-T2200419/public}.

The analysis reported in this paper also used the segment lists of
\cite{Goetz2023_Segments} and the self-gating procedure documented in
\cite{Zweizig2021_SelfGating}.

This research has made use of data available from the Gravitational
Wave Open Science Center (\url{http://gwosc.org}), a service of LIGO Laboratory,
the LIGO Scientific Collaboration, the Virgo Collaboration, and
KAGRA. LIGO Laboratory and Advanced LIGO are funded by the United
States National Science Foundation (NSF) as well as the Science and
Technology Facilities Council (STFC) of the United Kingdom, the
Max-Planck-Society (MPS), and the State of Niedersachsen/Germany for
support of the construction of Advanced LIGO and construction and
operation of the GEO600 detector. Additional support for Advanced LIGO
was provided by the Australian Research Council. Virgo is funded,
through the European Gravitational Observatory (EGO), by the French
Centre National de Recherche Scientifique (CNRS), the Italian Istituto
Nazionale di Fisica Nucleare (INFN) and the Dutch Nikhef, with
contributions by institutions from Belgium, Germany, Greece, Hungary,
Ireland, Japan, Monaco, Poland, Portugal, Spain. KAGRA is supported by
Ministry of Education, Culture, Sports, Science and Technology (MEXT),
Japan Society for the Promotion of Science (JSPS) in Japan; National
Research Foundation (NRF) and Ministry of Science and ICT (MSIT) in
Korea; Academia Sinica (AS) and National Science and Technology
Council (NSTC) in Taiwan.

\software{
\code{LALSuite}~\citep{LALSuite},
\code{LatticeTiling}~\citep{Wette2014_Lattice},
\code{PyFstat}~\citep{Ashton2018_PyFstat,Keitel2021_PyFstat,ashton_gregory_2022_7458002},
\code{ptemcee}~\citep{2013PASP..125..306F,2016MNRAS.455.1919V},
\code{numpy}~\citep{harris2020array},
\code{matplotlib}~\citep{matplotlib},
\code{scipy}~\citep{scipy},
\code{SwigLAL}~\citep{2020SoftX..1200634W}.
}

This paper has been assigned LIGO Document No.~\dcc.


\begin{thebibliography}{}
\expandafter\ifx\csname natexlab\endcsname\relax\def\natexlab#1{#1}\fi
\providecommand{\url}[1]{\href{#1}{#1}}
\providecommand{\dodoi}[1]{doi:~\href{http://doi.org/#1}{\nolinkurl{#1}}}
\providecommand{\doeprint}[1]{\href{http://ascl.net/#1}{\nolinkurl{http://ascl.net/#1}}}
\providecommand{\doarXiv}[1]{\href{https://arxiv.org/abs/#1}{\nolinkurl{https://arxiv.org/abs/#1}}}

\bibitem[{Aasi {et~al.}(2015{\natexlab{a}})Aasi, Abbott, Abbott,
  {et~al.}}]{LVC2015_S5ScoX1Sideband}
Aasi, J., Abbott, B.~P., Abbott, R., {et~al.} 2015{\natexlab{a}}, \prd, 91,
  062008, \dodoi{10.1103/PhysRevD.91.062008}

\bibitem[{Aasi {et~al.}(2015{\natexlab{b}})Aasi, Abbott, Abbott,
  {et~al.}}]{LVC2015_aLIGOInst}
---. 2015{\natexlab{b}}, \cqg, 32, 074001,
  \dodoi{10.1088/0264-9381/32/7/074001}

\bibitem[{Abbott {et~al.}(2017{\natexlab{a}})Abbott, Abbott, Abbott,
  {et~al.}}]{LVC2017_O1StochRadiometer}
Abbott, B.~P., Abbott, R., Abbott, T.~D., {et~al.} 2017{\natexlab{a}}, \prl,
  118, 121102, \dodoi{10.1103/PhysRevLett.118.121102}

\bibitem[{Abbott {et~al.}(2017{\natexlab{b}})Abbott, Abbott, Abbott,
  {et~al.}}]{LVC2017_O1ScoX1Viterbi}
---. 2017{\natexlab{b}}, \prd, 95, 122003, \dodoi{10.1103/PhysRevD.95.122003}

\bibitem[{Abbott {et~al.}(2017{\natexlab{c}})Abbott, Abbott, Abbott,
  {et~al.}}]{LVC2017_O1ScoX1CrossCorr}
---. 2017{\natexlab{c}}, \apj, 847, 47, \dodoi{10.3847/1538-4357/aa86f0}

\bibitem[{Abbott {et~al.}(2019{\natexlab{a}})Abbott, Abbott, Abbott,
  {et~al.}}]{LVC2019_O2StochRadiometer}
---. 2019{\natexlab{a}}, \prd, 100, 062001, \dodoi{10.1103/PhysRevD.100.062001}

\bibitem[{Abbott {et~al.}(2019{\natexlab{b}})Abbott, Abbott, Abbott,
  {et~al.}}]{LVC2019_O2ScoX1Viterbi}
---. 2019{\natexlab{b}}, \prd, 100, 122002, \dodoi{10.1103/PhysRevD.100.122002}

\bibitem[{Abbott {et~al.}(2021)Abbott, Abbott, Abraham,
  {et~al.}}]{LVK2021_O3StochRadiometer}
Abbott, B.~P., Abbott, T.~D., Abraham, S., {et~al.} 2021, \prd, 104, 022005,
  \dodoi{10.1103/PhysRevD.104.022005}

\bibitem[{Abbott {et~al.}(2022{\natexlab{a}})Abbott, Abe, Acernese,
  {et~al.}}]{LVK2022_O3ScoX1CrossCorr}
Abbott, B.~P., Abe, H., Acernese, F., {et~al.} 2022{\natexlab{a}}, \apjl, 941,
  L30, \dodoi{10.3847/2041-8213/aca1b0}

\bibitem[{Abbott {et~al.}(2022{\natexlab{b}})Abbott, Abe, Acernese,
  {et~al.}}]{LVK2022_O3ScoX1Viterbi}
---. 2022{\natexlab{b}}, \prd, 106, 062002, \dodoi{10.1103/PhysRevD.106.062002}

\bibitem[{{Abbott} {et~al.}(2023){Abbott}, {Abe}, {Acernese},
  {et~al.}}]{LVK2023_O3OpenData}
{Abbott}, R., {Abe}, H., {Acernese}, F., {et~al.} 2023, arXiv e-prints,
  arXiv:2302.03676.
\newblock \doarXiv{2302.03676}

\bibitem[{Abbott {et~al.}(2007)Abbott, Adhikari,
  {et~al.}}]{LSC2007_S2ScoX1FStat}
Abbott, B.~Abbott, R., Adhikari, R., {et~al.} 2007, \prd, 76, 082001,
  \dodoi{10.1103/PhysRevD.76.082001}

\bibitem[{Acernese {et~al.}(2015)Acernese, Agathos, Agatsuma,
  {et~al.}}]{Acernese2015_aVirgo}
Acernese, F., Agathos, M., Agatsuma, K., {et~al.} 2015, \cqg, 32, 024001,
  \dodoi{10.1088/0264-9381/32/2/024001}

\bibitem[{{Akutsu} {et~al.}(2021){Akutsu}, {Ando}, {Arai}, {Arai}, {Araki},
  {Araya}, {et~al.}}]{Akutsu2021_KAGRAO3}
{Akutsu}, T., {Ando}, M., {Arai}, K., {et~al.} 2021, Progress of Theoretical
  and Experimental Physics, 2021, 05A101, \dodoi{10.1093/ptep/ptaa125}

\bibitem[{Ashton {et~al.}(2022)Ashton, Keitel, Prix, \&
  Tenorio}]{ashton_gregory_2022_7458002}
Ashton, G., Keitel, D., Prix, R., \& Tenorio, R. 2022, PyFstat/PyFstat:
  v1.19.1, v1.19.1,  Zenodo, \dodoi{10.5281/zenodo.7458002}

\bibitem[{{Ashton} \& {Prix}(2018)}]{Ashton2018_PyFstat}
{Ashton}, G., \& {Prix}, R. 2018, \prd, 97, 103020,
  \dodoi{10.1103/PhysRevD.97.103020}

\bibitem[{Bildsten(1998)}]{Bildsten1998}
Bildsten, L. 1998, \apjl, 501, L89, \dodoi{10.1086/311440}

\bibitem[{Bradshaw {et~al.}(1999)Bradshaw, Fomalont, \&
  Geldzahler}]{Bradshaw1999_ScoX1}
Bradshaw, C.~F., Fomalont, E.~B., \& Geldzahler, B.~J. 1999, \apjl, 512, L121,
  \dodoi{10.1086/311889}

\bibitem[{Cutler \& Schutz(2005)}]{Cutler2005_FStat}
Cutler, C., \& Schutz, B.~F. 2005, \prd, 72, 063006,
  \dodoi{10.1103/PhysRevD.72.063006}

\bibitem[{{Davis} {et~al.}(2021){Davis}, {Areeda}, {Berger}, {Bruntz},
  {Effler}, {Essick}, {Fisher}, {Godwin}, {Goetz}, {Helmling-Cornell},
  {Hughey}, {Katsavounidis}, {Lundgren}, {Macleod}, {M{\'a}rka}, {Massinger},
  {Matas}, {McIver}, {Mo}, {Mogushi}, {Nguyen}, {Nuttall}, {Schofield},
  {Shoemaker}, {Soni}, {Stuver}, {Urban}, {Valdes}, {Walker}, {Abbott},
  {Adams}, {Adhikari}, {Ananyeva}, {Appert}, {Arai}, {Asali}, {Aston},
  {Austin}, {Baer}, {Ball}, {Ballmer}, {Banagiri}, {Barker}, {Barschaw},
  {Barsotti}, {Bartlett}, {Betzwieser}, {Beda}, {Bhattacharjee}, {Bidler},
  {Billingsley}, {Biscans}, {Blair}, {Blair}, {Bode}, {Booker}, {Bork},
  {Bramley}, {Brooks}, {Brown}, {Buikema}, {Cahillane}, {Callister}, {Caneva
  Santoro}, {Cannon}, {Carlin}, {Chandra}, {Chen}, {Christensen}, {Ciobanu},
  {Clara}, {Compton}, {Cooper}, {Corley}, {Coughlin}, {Countryman}, {Covas},
  {Coyne}, {Crowder}, {Dal Canton}, {Danila}, {Datrier}, {Davies}, {Dent},
  {Didio}, {Di Fronzo}, {Dooley}, {Driggers}, {Dupej}, {Dwyer}, {Etzel},
  {Evans}, {Evans}, {Fairhurst}, {Feicht}, {Fernandez-Galiana}, {Frey},
  {Fritschel}, {Frolov}, {Fulda}, {Fyffe}, {Gadre}, {Giaime}, {Giardina},
  {Gonz{\'a}lez}, {Gras}, {Gray}, {Gray}, {Green}, {Gupta}, {Gustafson},
  {Gustafson}, {Hanks}, {Hanson}, {Hardwick}, {Harry}, {Hasskew}, {Heintze},
  {Heinzel}, {Holland}, {Hollows}, {Hoy}, {Hughey}, {Jadhav}, {Janssens},
  {Johns}, {Jones}, {Kandhasamy}, {Karki}, {Kasprzack}, {Kawabe}, {Keitel},
  {Kijbunchoo}, {Kim}, {King}, {Kissel}, {Kulkarni}, {Kumar}, {Landry}, {Lane},
  {Lantz}, {Laxen}, {Lecoeuche}, {Leviton}, {Liu}, {Lormand}, {Macas},
  {Macedo}, {MacInnis}, {Mandic}, {Mansell}, {M{\'a}rka}, {Martinez},
  {Martinovic}, {Martynov}, {Mason}, {Matichard}, {Mavalvala}, {McCarthy},
  {McClelland}, {McCormick}, {McCuller}, {McIsaac}, {McRae}, {Mendell},
  {Merfeld}, {Merilh}, {Meyers}, {Meylahn}, {Michaloliakos}, {Middleton},
  {Mills}, {Mistry}, {Mittleman}, {Moreno}, {Mow-Lowry}, {Mozzon}, {Mueller},
  {Mukund}, {Mullavey}, {Muth}, {Nelson}, {Neunzert}, {Nichols}, {Nitoglia},
  {Oberling}, {Oh}, {Oh}, {Oram}, {Ormiston}, {Ormsby}, {Osthelder}, {Ottaway},
  {Overmier}, {Pai}, {Palamos}, {Pannarale}, {Parker}, {Patane}, {Patel},
  {Payne}, {Pele}, {Penhorwood}, {Perez}, {Phukon}, {Pillas}, {Pirello},
  {Radkins}, {Ramirez}, {Richardson}, {Riles}, {Rink}, {Robertson}, {Rollins},
  {Romel}, {Romie}, {Ross}, {Ryan}, {Sadecki}, {Sakellariadou}, {Sanchez},
  {Sanchez}, {Sandles}, {Saravanan}, {Savage}, {Schaetzl}, {Schnabel},
  {Schwartz}, {Sellers}, {Shaffer}, {Sigg}, {Sintes}, {Slagmolen}, {Smith},
  {Soni}, {Sorazu}, {Spencer}, {Strain}, {Strom}, {Sun}, {Szczepa{\'n}czyk},
  {Tasson}, {Tenorio}, {Thomas}, {Thomas}, {Thorne}, {Toland}, {Torrie},
  {Tran}, {Traylor}, {Trevor}, {Tse}, {Vajente}, {van Remortel}, {Vander-Hyde},
  {Vargas}, {Veitch}, {Veitch}, {Venkateswara}, {Venugopalan}, {Viets},
  {Villa-Ortega}, {Vo}, {Vorvick}, {Wade}, {Wallace}, {Ward}, {Warner},
  {Weaver}, {Weinstein}, {Weiss}, {Wette}, {White}, {White}, {Whittle},
  {Williamson}, {Willke}, {Wipf}, {Xiao}, {Xu}, {Yamamoto}, {Yu}, {Yu},
  {Zhang}, {Zheng}, {Zucker}, \& {Zweizig}}]{Davis2021_DetChar}
{Davis}, D., {Areeda}, J.~S., {Berger}, B.~K., {et~al.} 2021, \cqg, 38, 135014,
  \dodoi{10.1088/1361-6382/abfd85}

\bibitem[{Dhurandhar {et~al.}(2008)Dhurandhar, Krishnan, Mukhopadhyay, \&
  Whelan}]{Dhurandhar2007_CrossCorr}
Dhurandhar, S., Krishnan, B., Mukhopadhyay, H., \& Whelan, J.~T. 2008, \prd,
  77, 082001, \dodoi{10.1103/PhysRevD.77.082001}

\bibitem[{{Foreman-Mackey} {et~al.}(2013){Foreman-Mackey}, {Hogg}, {Lang}, \&
  {Goodman}}]{2013PASP..125..306F}
{Foreman-Mackey}, D., {Hogg}, D.~W., {Lang}, D., \& {Goodman}, J. 2013, \pasp,
  125, 306, \dodoi{10.1086/670067}

\bibitem[{{Galloway} {et~al.}(2014){Galloway}, {Premachandra}, {Steeghs},
  {Marsh}, {Casares}, \& {Cornelisse}}]{Galloway2014_PEGS1}
{Galloway}, D.~K., {Premachandra}, S., {Steeghs}, D., {et~al.} 2014, \apj, 781,
  14, \dodoi{10.1088/0004-637X/781/1/14}

\bibitem[{Goetz {et~al.}(2021)Goetz, Neunzert, Riles,
  {et~al.}}]{Goetz2021_Lines}
Goetz, E., Neunzert, A., Riles, K., {et~al.} 2021, {O3 lines and combs in found
  in self-gated C01 data}, LIGO Document T2100200-v2.
\newblock \url{https://dcc.ligo.org/LIGO-T2100200/public}

\bibitem[{Goetz \& Riles(2023)}]{Goetz2023_Segments}
Goetz, E., \& Riles, K. 2023, {Segments used for creating standard SFTs in O3
  data}, LIGO Document T2300068-v2.
\newblock \url{https://dcc.ligo.org/LIGO-T2300068/public}

\bibitem[{Harris {et~al.}(2020)Harris, Millman, van~der Walt,
  {et~al.}}]{harris2020array}
Harris, C.~R., Millman, K.~J., van~der Walt, S.~J., {et~al.} 2020, Nature, 585,
  357, \dodoi{10.1038/s41586-020-2649-2}

\bibitem[{{Hunter}(2007)}]{matplotlib}
{Hunter}, J.~D. 2007, CSE, 9, 90, \dodoi{10.1109/MCSE.2007.55}

\bibitem[{Jaranowski {et~al.}(1998)Jaranowski, Krolak, \&
  Schutz}]{JKS1998_FStat}
Jaranowski, P., Krolak, A., \& Schutz, B.~F. 1998, \prd, 58, 063001,
  \dodoi{10.1103/PhysRevD.58.063001}

\bibitem[{{Keitel} {et~al.}(2021){Keitel}, {Tenorio}, {Ashton}, \&
  {Prix}}]{Keitel2021_PyFstat}
{Keitel}, D., {Tenorio}, R., {Ashton}, G., \& {Prix}, R. 2021, The Journal of
  Open Source Software, 6, 3000, \dodoi{10.21105/joss.03000}

\bibitem[{{Killestein} {et~al.}(2023){Killestein}, {Mould}, {Steeghs},
  {Casares}, \& {Galloway}}]{Killestein2023_PEGS4}
{Killestein}, T.~L., {Mould}, M., {Steeghs}, D., {Casares}, J., \& {Galloway},
  D.~K. 2023, \mnras, \dodoi{10.1093/mnras/stad366}

\bibitem[{{LIGO Scientific Collaboration}(2018)}]{LALSuite}
{LIGO Scientific Collaboration}. 2018, {LIGO} {A}lgorithm {L}ibrary -
  {LALS}uite, free software (GPL), \dodoi{10.7935/GT1W-FZ16}

\bibitem[{{LIGO-Virgo-KAGRA Collaboration}(2023)}]{LVK2023_ObsPlanUpdate}
{LIGO-Virgo-KAGRA Collaboration}. 2023, LIGO, Virgo and KAGRA Observing Run
  Plans (19 January 2023 Update, https://observing.docs.ligo.org/plan/.
\newblock \url{https://observing.docs.ligo.org/plan/}

\bibitem[{Meadors {et~al.}(2017)Meadors, Goetz, Riles, Creighton, \&
  Robinet}]{Meadors2017_S6ScoX1TwoSpect}
Meadors, G.~D., Goetz, E., Riles, K., Creighton, T., \& Robinet, F. 2017, \prd,
  95, 042005, \dodoi{10.1103/PhysRevD.95.042005}

\bibitem[{{OSG}(2006)}]{OSGPool}
{OSG}. 2006, OSPool, computing resource, \dodoi{10.21231/906P-4D78}

\bibitem[{Pordes {et~al.}(2007)Pordes, Petravick, Kramer, Olson, Livny, Roy,
  Avery, Blackburn, Wenaus, W{\"u}rthwein, Foster, Gardner, Wilde, Blatecky,
  McGee, \& Quick}]{Pordes2007_OSG}
Pordes, R., Petravick, D., Kramer, B., {et~al.} 2007, in 78, Vol.~78, J. Phys.
  Conf. Ser., 012057, \dodoi{10.1088/1742-6596/78/1/012057}

\bibitem[{{Premachandra} {et~al.}(2016){Premachandra}, {Galloway}, {Casares},
  {Steeghs}, \& {Marsh}}]{Premachandra2016_PEGS2}
{Premachandra}, S.~S., {Galloway}, D.~K., {Casares}, J., {Steeghs}, D.~T., \&
  {Marsh}, T.~R. 2016, \apj, 823, 106, \dodoi{10.3847/0004-637X/823/2/106}

\bibitem[{Sfiligoi {et~al.}(2009)Sfiligoi, Bradley, Holzman, Mhashilkar, Padhi,
  \& Wurthwein}]{Sfiligoi2009_OSG}
Sfiligoi, I., Bradley, D.~C., Holzman, B., {et~al.} 2009, in 2, Vol.~2, 2009
  WRI World Congress on Computer Science and Information Engineering, 428--432,
  \dodoi{10.1109/CSIE.2009.950}

\bibitem[{{Steeghs} \& {Casares}(2002)}]{Steeghs2002_ScoX1}
{Steeghs}, D., \& {Casares}, J. 2002, \apj, 568, 273, \dodoi{10.1086/339224}

\bibitem[{{Suvorova} {et~al.}(2017){Suvorova}, {Clearwater}, {Melatos}, {Sun},
  {Moran}, \& {Evans}}]{Suvorova2017_Viterbi2}
{Suvorova}, S., {Clearwater}, P., {Melatos}, A., {et~al.} 2017, \prd, 96,
  102006, \dodoi{10.1103/PhysRevD.96.102006}

\bibitem[{Suvorova {et~al.}(2016)Suvorova, Sun, Melatos, Moran, \&
  Evans}]{Suvorova2016_Viterbi}
Suvorova, S., Sun, L., Melatos, A., Moran, W., \& Evans, R. 2016, \prd, 93,
  123009, \dodoi{10.1103/PhysRevD.93.123009}

\bibitem[{{Tenorio} {et~al.}(2021){Tenorio}, {Keitel}, \&
  {Sintes}}]{Tenorio2021_MCMC}
{Tenorio}, R., {Keitel}, D., \& {Sintes}, A.~M. 2021, \prd, 104, 084012,
  \dodoi{10.1103/PhysRevD.104.084012}

\bibitem[{Virtanen {et~al.}(2020)Virtanen, Gommers, Oliphant, {et~al.}}]{scipy}
Virtanen, P., Gommers, R., Oliphant, T.~E., {et~al.} 2020, Nature Methods, 17,
  261, \dodoi{10.1038/s41592-019-0686-2}

\bibitem[{{Vousden} {et~al.}(2016){Vousden}, {Farr}, \&
  {Mandel}}]{2016MNRAS.455.1919V}
{Vousden}, W.~D., {Farr}, W.~M., \& {Mandel}, I. 2016, \mnras, 455, 1919,
  \dodoi{10.1093/mnras/stv2422}

\bibitem[{Wagner {et~al.}(2022)Wagner, Whelan, Wofford, \&
  Wette}]{Wagner2022_Lattice}
Wagner, K.~J., Whelan, J.~T., Wofford, J.~K., \& Wette, K. 2022, \cqg, 39,
  075013, \dodoi{10.1088/1361-6382/ac5012}

\bibitem[{{Wang} {et~al.}(2018){Wang}, {Steeghs}, {Galloway}, {Marsh}, \&
  {Casares}}]{Wang2018_PEGS3}
{Wang}, L., {Steeghs}, D., {Galloway}, D.~K., {Marsh}, T., \& {Casares}, J.
  2018, \mnras, 478, 5174, \dodoi{10.1093/mnras/sty1441}

\bibitem[{Watts {et~al.}(2008)Watts, Krishnan, Bildsten, \& Schutz}]{Watts2008}
Watts, A.~L., Krishnan, B., Bildsten, L., \& Schutz, B.~F. 2008, \mnras, 389,
  839

\bibitem[{Wette(2014)}]{Wette2014_Lattice}
Wette, K. 2014, \prd, 90, 122010, \dodoi{10.1103/PhysRevD.90.122010}

\bibitem[{{Wette}(2020)}]{2020SoftX..1200634W}
{Wette}, K. 2020, SoftwareX, 12, 100634, \dodoi{10.1016/j.softx.2020.100634}

\bibitem[{Whelan {et~al.}(2015)Whelan, Sundaresan, Zhang, \&
  Peiris}]{Whelan2015_ScoX1CrossCorr}
Whelan, J.~T., Sundaresan, S., Zhang, Y., \& Peiris, P. 2015, \prd, 91, 102005,
  \dodoi{10.1103/PhysRevD.91.102005}

\bibitem[{{Zhang} {et~al.}(2021){Zhang}, {Papa}, {Krishnan}, \&
  {Watts}}]{Zhang2021_O2ScoX1CrossCorr}
{Zhang}, Y., {Papa}, M.~A., {Krishnan}, B., \& {Watts}, A.~L. 2021, \apjl, 906,
  L14, \dodoi{10.3847/2041-8213/abd256}

\bibitem[{Zweizig \& Riles(2021)}]{Zweizig2021_SelfGating}
Zweizig, J., \& Riles, K. 2021, {Information on self-gating of $h(t)$ used in
  O3 continuous-wave and stochastic searches}, LIGO Document T2000384-v4.
\newblock \url{https://dcc.ligo.org/LIGO-T2000384/public}

\end{thebibliography}
\end{document}